\documentclass[a4paper,twocolumn,amsmath,amssymb,superscriptaddress,nofootinbib,preprintnumbers]{revtex4}

\usepackage{textcomp}
\usepackage{dcolumn}
\usepackage{amsmath}
\usepackage{graphicx}
\usepackage{rotating}
\usepackage{amsfonts}
\usepackage{amssymb}
\usepackage[hang]{subfigure}

\usepackage{graphicx}
\usepackage{epsfig}

\newcommand{\lwig}{\mbox{\,\raisebox{.3ex}
    {$<$}$\!\!\!\!\!$\raisebox{-.9ex}{$\sim$}\,}}

\def\DESY{Deutsches Elektronen-Synchrotron DESY, Notkestra\ss e 85, D-22607 Hamburg, Germany}
\def\LZH{Laser Zentrum Hannover e.V., Hollerithallee 8, D-30419 Hannover, Germany}
\def\Sternwarte{Sternwarte Bergedorf, Gojenbergsweg 112, D-21029 Hamburg, Germany}

\begin{document}

\preprint{DESY 07-014\hspace{12.cm}  Public version --- 8.12.2006}

\title {
Production and Detection of Axion-Like Particles in a HERA Dipole Magnet\\[1ex]
-- Letter-of-Intent for the ALPS experiment --}

\author{Klaus Ehret}
\affiliation{\DESY}
\author{Maik Frede}
\affiliation{\LZH}
\author{Ernst-Axel Knabbe}
\affiliation{\DESY}
\author{Dietmar Kracht}
\affiliation{\LZH}
\author{Axel Lindner}
\email{axel.lindner@desy.de}
\affiliation{\DESY}
\author{Niels Meyer}
\affiliation{\DESY}
\author{Dieter Notz}
\affiliation{\DESY}
\author{Andreas Ringwald}
\email{andreas.ringwald@desy.de}
\affiliation{\DESY}
\author{G\"unter Wiedemann}
\affiliation{\Sternwarte}

\begin{abstract}
Recently, the PVLAS collaboration has reported evidence for an 
anomalous rotation of the polarization 
of light in vacuum in the presence of a transverse magnetic field. 
This may be explained  through the production of a 
new light spin-zero (axion-like) neutral particle coupled to two photons. 
In this letter-of-intent, we propose to test 
this hypothesis by setting up a photon regeneration experiment
which exploits the photon beam of a high-power infrared laser, 
sent along the transverse magnetic field of a superconducting HERA dipole 
magnet. The proposed\footnote{The experiment has been approved by the DESY 
directorate on January 11, 2007.} ALPS (Axion-Like 
Particle Search) experiment offers a window of opportunity for a rapid
firm establishment or exclusion of the axion-like particle interpretation 
of the anomaly published by PVALS. It will also allow for the measurement of mass, 
parity, and coupling strength of this particle.
\end{abstract}


\maketitle

\section{Introduction and Motivation}

New very light spin-zero particles are predicted in many models beyond the Standard Model. Often they 
would couple only very weakly to ordinary matter. Typically, such light particles arise  
if there is a global continuous symmetry in the theory that is spontaneously broken in the vacuum
--- a notable example being the axion~\cite{Weinberg:1978ma}, a pseudo scalar particle 
 arising from the breaking of a U(1) 
Peccei-Quinn symmetry~\cite{Peccei:1977hh}, 
introduced to explain the absence of $CP$ violation in strong interactions.
Such axion-like pseudo scalars couple to two photons via 
\begin{equation}
{\cal L}_{\phi \gamma \gamma} = - \frac{1}{4}\, g\, \phi\, F_{\mu \nu}\tilde{F}^{\mu \nu} = 
g\, \phi\, \vec{E}\cdot \vec{B} , 
\label{coupling}
\end{equation}
where $g$ is the coupling, $\phi$ is the field corresponding to the particle, 
$F_{\mu\nu}$ ($\tilde{F}^{\mu\nu}$) is the (dual) electromagnetic field strength tensor,
and $\vec{E}$ and $\vec{B}$ are the electric and magnetic fields, respectively. 
\begin{figure}[t]
\begin{center}
\includegraphics*[bbllx=86pt,bblly=637pt,bburx=298pt,bbury=707pt,width=8.cm,clip=]{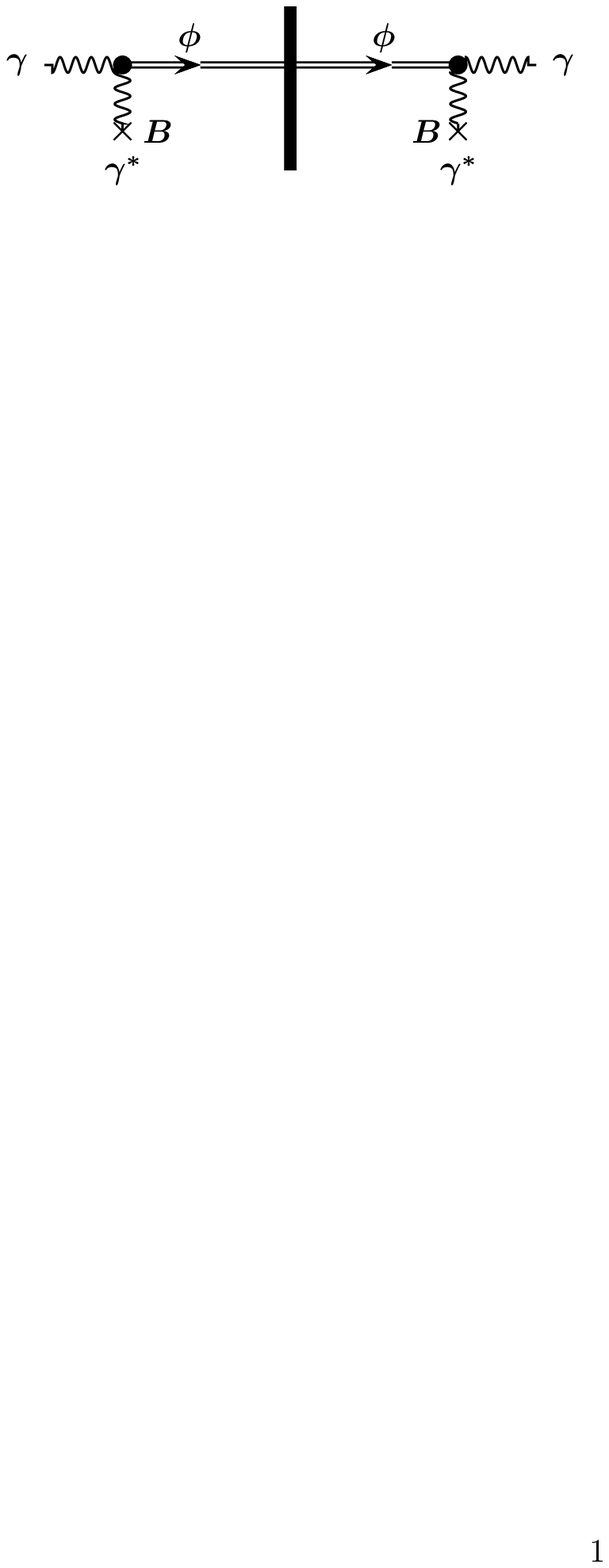}
\caption[...]{Schematic view of (pseudo-)scalar production through photon conversion 
in a magnetic field (left), subsequent travel through a wall, and final 
detection through photon regeneration (right). 
\hfill
\label{fig:ph_reg}}
\end{center}
\end{figure}
In the case of a scalar particle coupling to two photons, the interaction 
reads
\begin{equation}
{\cal L}_{\phi \gamma \gamma} =  -\frac{1}{4}\, g\, \phi\, F_{\mu \nu} F^{\mu \nu} = 
\frac{1}{2}g\, \phi\, \left( \vec{E}^2 - \vec{B}^2 \right).  
\label{coupling_scalar}
\end{equation}
Both effective interactions give rise to similar observable effects. In particular, 
in the presence of an external magnetic field, a photon of frequency $\omega$ may oscillate into a 
light spin-zero particle  of small mass $m_\phi < \omega$, and vice versa.  
The notable difference between a pseudo scalar and a scalar  is that it is
the component of the photon polarization parallel to the magnetic field that interacts in
the former case, whereas it is the perpendicular component in the latter case. 

\begin{figure}[t]
\begin{center}
\includegraphics*[bbllx=25,bblly=226,bburx=584,bbury=604,width=.45\textwidth]{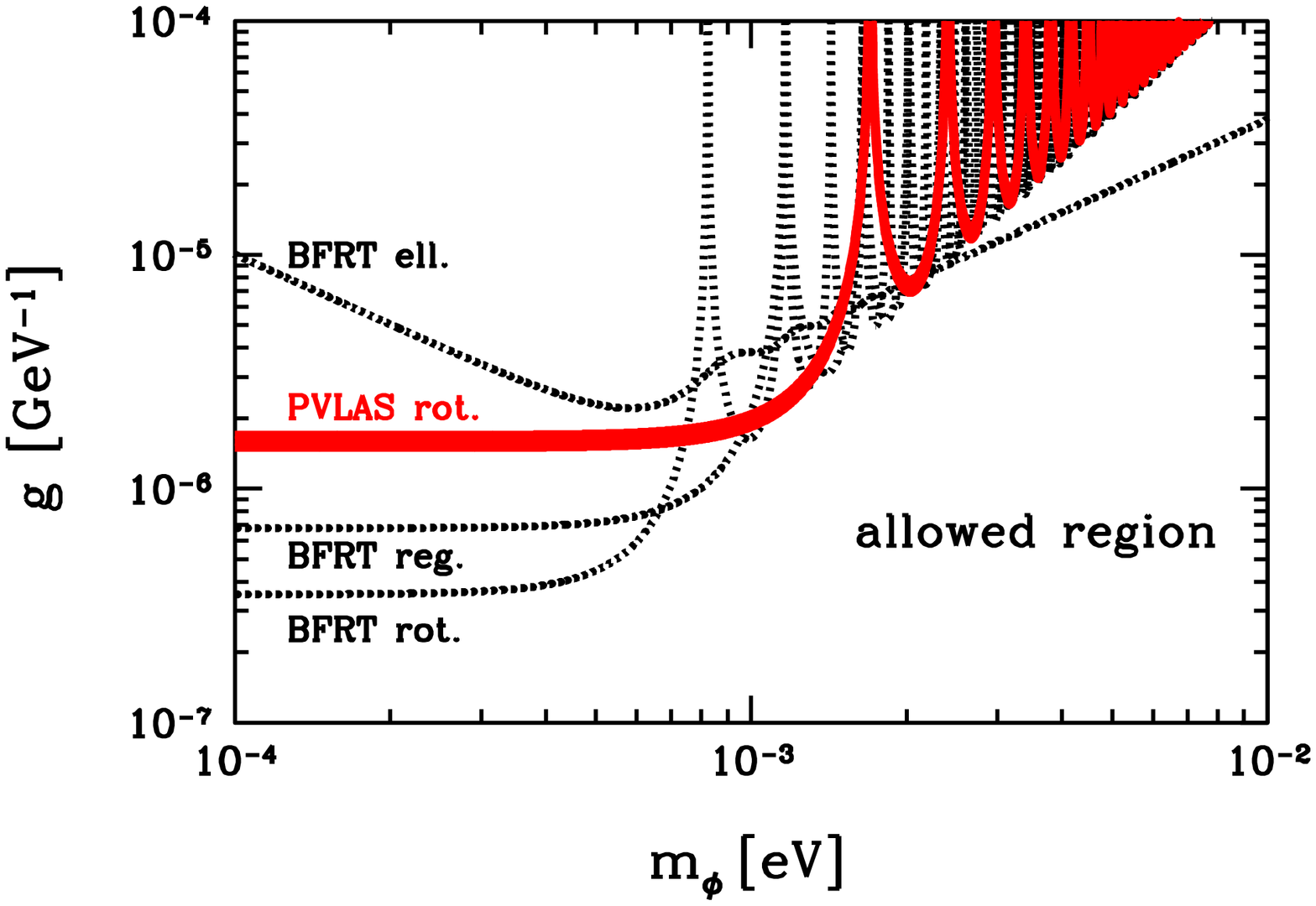}

\vspace{2ex}
\includegraphics*[bbllx=25,bblly=226,bburx=584,bbury=604,width=.45\textwidth]{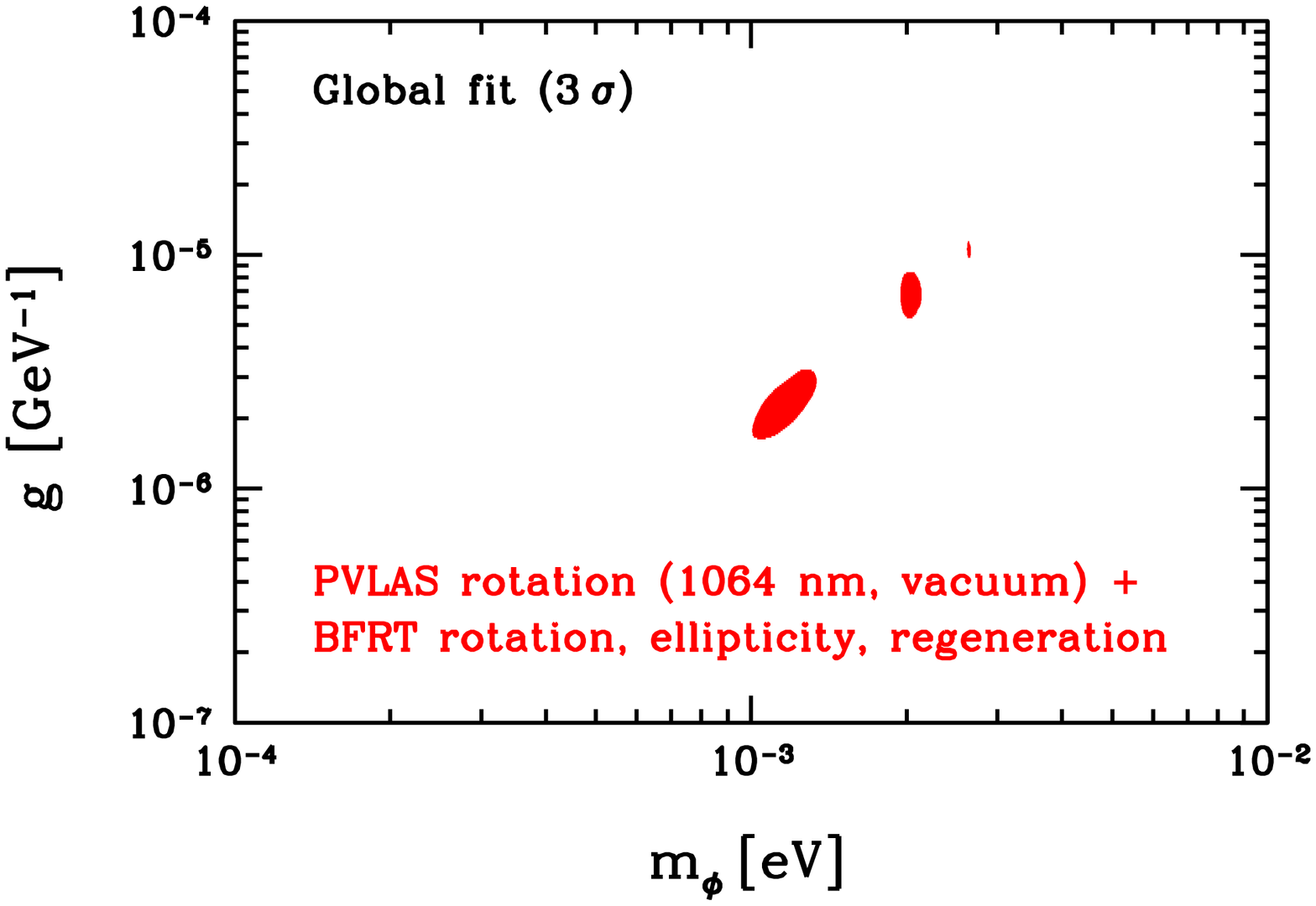}
\caption[...]{Two photon coupling $g$ of the (pseudo-)scalar versus its mass $m_\phi$.\\
Top panel: 
The 95\,\% confidence level upper limits from BFRT data~\cite{Cameron:1993mr}  
on polarization (rotation and ellipticity data)  
and photon regeneration are displayed as dotted lines.. 
The preferred values corresponding to the anomalous rotation signal observed by PVLAS~\cite{Zavattini:2005tm}  
are shown as a thick solid line.\\
Bottom panel: 
Three sigma allowed region from PVLAS data on rotation plus BFRT data on rotation, 
ellipticity, and regeneration~\cite{Ahlers:priv}.  
\hfill
\label{fig:ax_ph_lab_only}}
\end{center}
\end{figure}

The exploitation of this mechanism is the basic idea behind photon regeneration 
(sometimes called ``light shining through walls'') 
experiments~\cite{regeneration,VanBibber:1987rq}, see Fig.~\ref{fig:ph_reg}. Namely, if a beam of  
photons is shone across a magnetic field, a fraction of these photons will turn into (pseudo-)scalars.  
This (pseudo-)scalar beam could then propagate freely through a wall or another obstruction without being 
absorbed,  
and finally another magnetic field located on the other side of the wall could transform some of these 
(pseudo-)scalars 
into photons --- apparently regenerating these photons out of nothing. 
A pilot experiment of this type was carried out in Brookhaven using two prototype magnets for the 
Colliding Beam Accelerator~\cite{Ruoso:1992nx}. 
From the non-observation of photon regeneration, the Brookhaven-Fermilab-Rochester-Trieste
(BFRT) collaboration excluded values of the  coupling 
$g< 6.7  \times 10^{-7}\ {\rm GeV}^{-1}$, for $m_\phi \lwig 10^{-3}$ {\rm eV}~\cite{Cameron:1993mr} 
(cf. Fig.~\ref{fig:ax_ph_lab_only}), at the 95\,\% confidence level.   

Recently, the PVLAS collaboration has reported an anomalous signal in 
measurements of the rotation of the polarization plane of a laser beam in a magnetic 
field~\cite{Zavattini:2005tm}. 
A possible explanation of such an apparent vacuum magnetic dichroism is through the 
production of a light pseudo scalar or scalar, coupled to photons through Eq.~(\ref{coupling}) or 
Eq.~(\ref{coupling_scalar}), respectively.  Accordingly, 
photons polarized parallel (pseudo scalar) 
or perpendicular (scalar) 
to the magnetic field disappear, leading to a rotation of the 
polarization plane~\cite{Maiani:1986md}. The region quoted in Ref.~\cite{Zavattini:2005tm} that might explain 
the observed signal is around (95\,\% confidence level)
\begin{eqnarray}
\label{PVLAS_coupling}
1.7 \times 10^{-6}\  {\rm GeV}^{-1}< & g & < 5.0 \times 10^{-6}\  {\rm GeV}^{-1},
\\[2ex]
\label{PVLAS_mass}
1.0 \times  10^{-3}\  {\rm eV} < & m_\phi & < 1.5 \times  10^{-3}\  {\rm eV},
\end{eqnarray}
obtained from a combination of
previous limits on $g$ vs. $m_\phi$ from a similar, but less sensitive polarization experiment 
performed by the BFRT collaboration~\cite{Cameron:1993mr} and the $g$ vs. $m_\phi$ curve 
corresponding to the PVLAS signal (cf. Fig.~\ref{fig:ax_ph_lab_only} (lower panel) where also two other 
small allowed (at 3 sigma) regions  are displayed.).   

A particle with these properties presents a theoretical challenge. It is hardly compatible with a 
genuine QCD axion~\cite{Raffelt:2005mt,Ringwald:2005gf}.  
Moreover, it must have very peculiar properties in order to evade the 
strong constraints on $g$ from stellar energy loss considerations~\cite{Raffelt:1985nk}
and from its non-observation 
in the CERN Axion Solar Telescope~\cite{Andriamonje:2004hi,Raffelt:2005mt}. 
Its production in stars may be hindered, for example, 
if the $\phi\gamma\gamma$ vertex is suppressed at keV energies due to low scale 
compositeness of $\phi$, or if, in stellar interiors, 
$\phi$ acquires an effective mass larger than the typical photon energy, $\sim$~keV,  
or if the particles are trapped within stars~\cite{Masso:2005ym,Jain:2005nh,Jaeckel:2006id,Mohapatra:2006pv}.  
   
Clearly, an independent and decisive experimental test of the particle interpretation of
the PVLAS observation, without reference to axion production in stars 
(see~\cite{Dupays:2005xs,Kleban:2005rj,Fairbairn:2006hv}), is urgently needed~\cite{Ahlers:2006iz}. 
In Ref.~\cite{Ringwald:2003ns}, one of us (AR) proposed to exploit the strong magnetic field
of superconducting HERA dipole magnets for a photon regeneration experiment. 
In this letter-of-intent, we propose a possible realization of such an experiment.

Within the international community the PVLAS results have triggered
substantial activities. Besides this proposal efforts are under way in
Europe (at CERN, INFN and within a collaboration of different laboratories
in France) and the US (at Jefferson Laboratories) to directly test the
particle interpretation of the PVLAS result (for a review
see \cite{Ringwald:2006rf}). Recently a corresponding
workshop took place at the Institute for Advanced Studies~\cite{workshop} in Princeton. 
According to the presented
time schedules ALPS could be the first experiment to clarify
experimentally whether a previously unknown particle can be created by
interactions of photons with a strong magnetic field.
 
\begin{figure}
\begin{center}
\includegraphics*[bbllx=37,bblly=239,bburx=596,bbury=621,width=.45\textwidth]{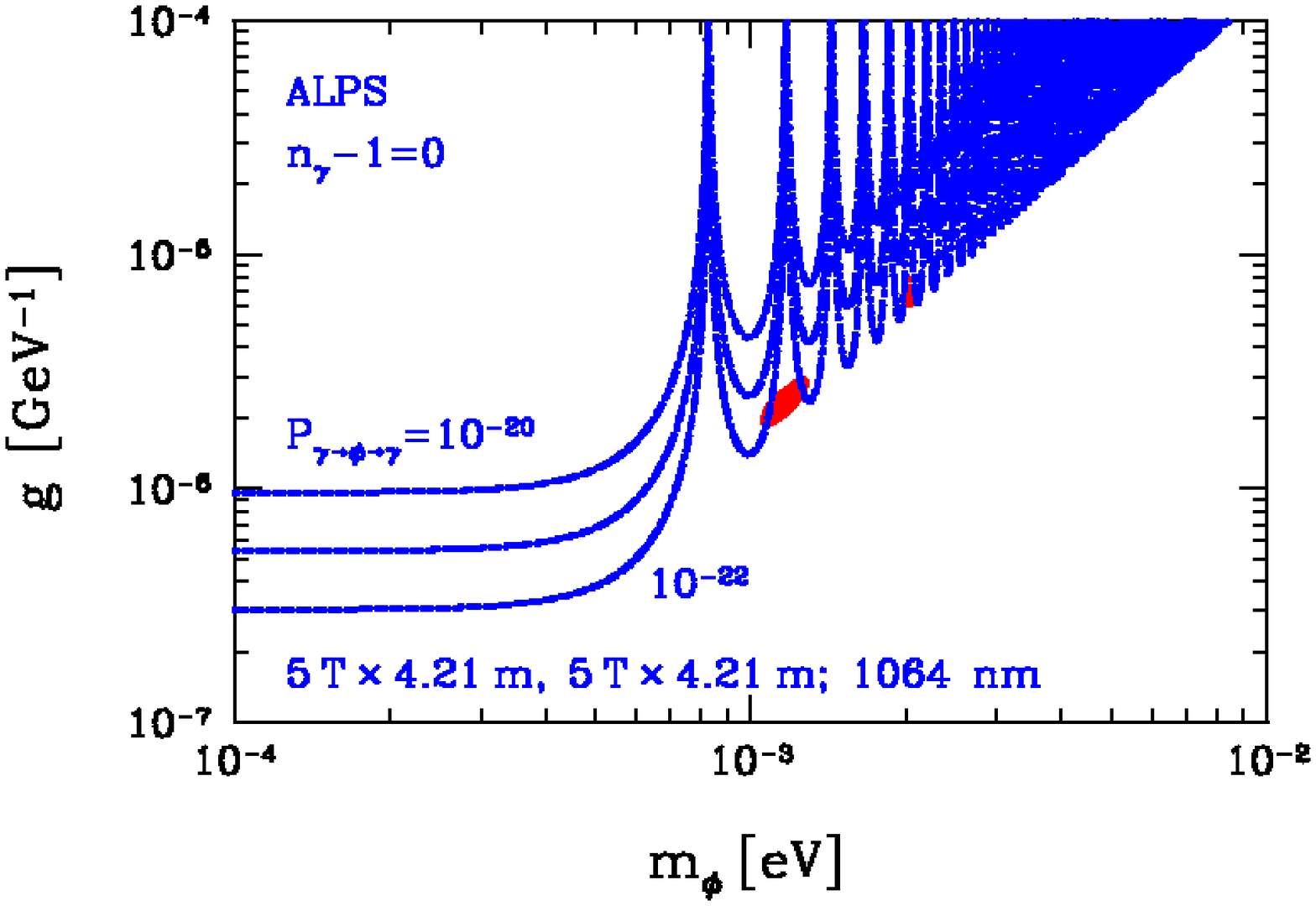}
\includegraphics*[bbllx=26,bblly=222,bburx=583,bbury=610,width=.45\textwidth]{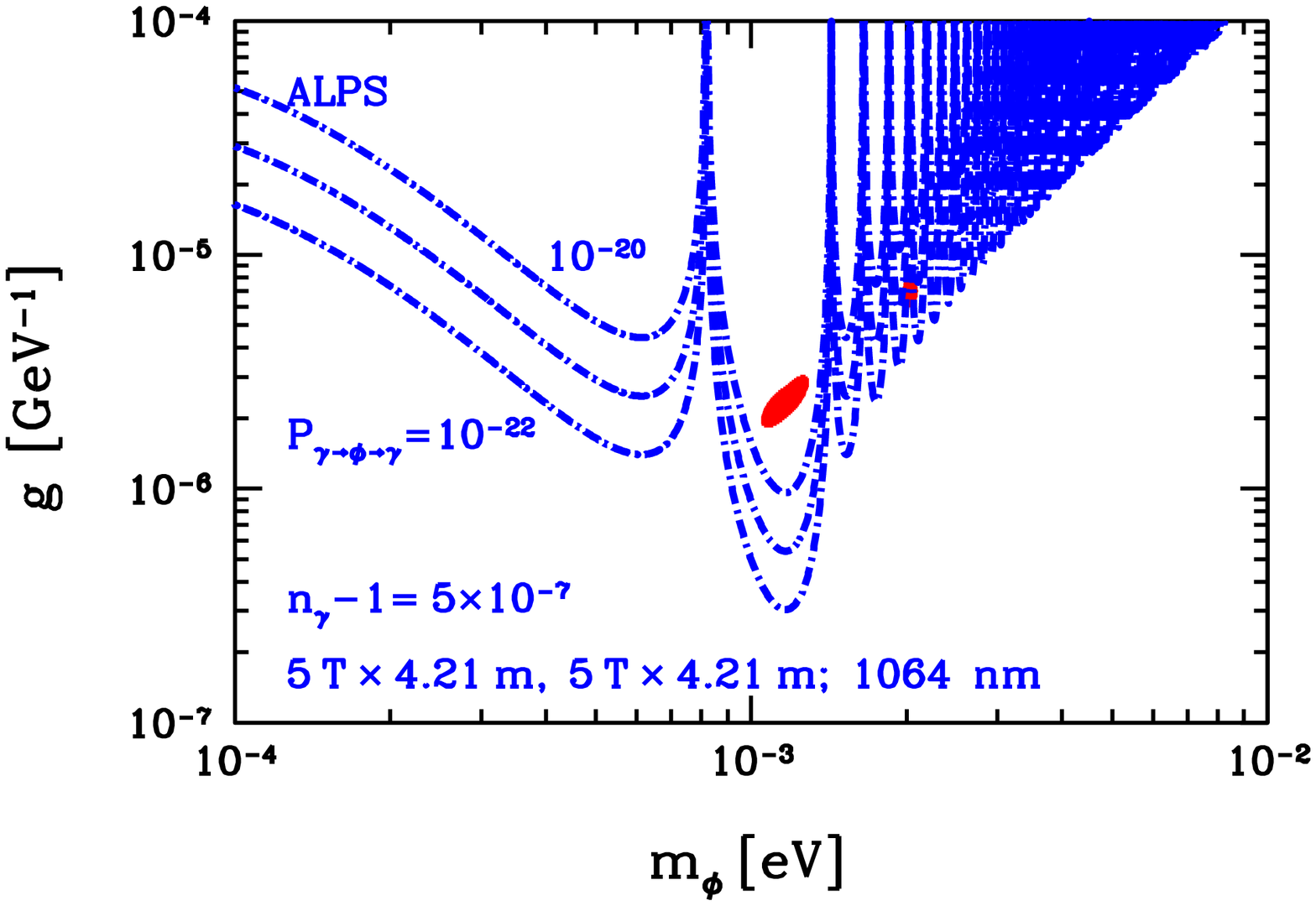}
\caption[...]{Two photon coupling $g$ of the (pseudo-)scalar versus its mass $m_\phi$. 
The 3 sigma allowed region from PVLAS data on rotation plus BFRT data on rotation, 
ellipticity, and regeneration is shown in red~\cite{Ahlers:priv}.\\  
Top panel: Iso-contours of the regeneration probability probability $P_{\gamma\to\phi\to\gamma}$, 
Eq.~(\ref{regpropab}), for the parameters of the HERA dipole magnet, $B_1=B_2=5$~T, 
$\ell_1=\ell_2=4.21$~m, exploiting an infrared photon beam, $\lambda =1064$~nm, corresponding
to $\omega =1.17$~eV, in vacuum ($n_\gamma =1$).\\
Bottom panel: Same as top panel, but in buffer gas with $n_\gamma -1=5\times 10^{-7}$.
\hfill
\label{regprop}}
\end{center}
\end{figure}

\section{Photon regeneration in a HERA dipole magnet}

We start with a discussion of the probability $P_{\gamma\to\phi\to\gamma}$ that an 
initial photon with energy $\omega$ converts, 
in the magnetic field region of size $B_1$ and length $\ell_1$ in front of the wall, 
into an axion-like particle, 
and reconverts, in the magnetic field region of size $B_2$ and length $\ell_2$ on the other side of the wall, 
into a photon. It is given by
\begin{equation}
\label{regpropab}
P_{\gamma\to\phi\to\gamma}=P_{\gamma\to\phi}(B_1,\ell_1,q_1)\,P_{\phi\to\gamma}(B_2,\ell_2,q_2)\,,
\end{equation}
where $P_{\gamma\to\phi}\equiv P_{\phi\to\gamma}$ is the probability that a photon converts into 
an axion-like particle,
\begin{equation}
P_{\gamma\to\phi}(B,\ell,q) = \frac{1}{4} \left(g\,B\,\ell\right)^2\,F(q\ell)\,.
\end{equation}
Here, 
\begin{equation}
q=\left|\frac{m_\gamma^2-m_\phi^2}{2\omega }\right|\hspace{3ex} (\ll m_\phi) 
\end{equation}
is the momentum transfer to the magnetic field, i.e. the modulus of the 
momentum difference between the photon and 
the axion-like particle, and 
\begin{equation}
\label{formfactor}
F(q\ell ) = \left[\frac{\sin\left( \frac{1}{2}q\ell\right)}{\frac{1}{2}q\ell}\right]^2
\end{equation}
is a form factor which reduces to unity for small momentum transfer $q\ell\ll 1$. 
For large $q\ell$, incoherence effects emerge between the (pseudo-)scalar and the 
photon, the form factor getting much smaller than unity, 
severely reducing the conversion probability. Clearly, in vacuum, the photon mass vanishes, 
$m_\gamma =0$. In a refractive medium, which may be realized in our experiment 
by filling in buffer gas~\cite{vanBibber:1988ge}, the effective mass is given by
\begin{equation}
m_\gamma^2 = 2\,(n_\gamma -1)\,\omega^2\,,
\end{equation}
where $n_\gamma$ is the refraction index of the medium. 
Therefore, by tuning $n_\gamma$, i.e. by varying the gas pressure in the magnetic field regions, 
one may optimize the sensitivity in certain mass regions by essentially
tuning $q$ toward small values.

In Fig.~\ref{regprop}, we display iso-contours of the regeneration probability 
$P_{\gamma\to\phi\to\gamma}$ in the $g$--$m_\phi$--plane, for the parameters 
of the HERA dipole magnet, $B_1=B_2=5$~T, 
$\ell_1=\ell_2=4.21$~m, exploiting an infrared photon beam, $\lambda =1064$~nm, corresponding
to $\omega =1.17$~eV, in vacuum ($n_\gamma =1$; top panel) and with buffer gas 
($n_\gamma -1=5\times 10^{-7}$; bottom panel). We infer from that figure that it will be very 
important to have the possibility to run the experiment with a buffer gas of 
an optimal refraction index.
In order to probe the PVLAS preferred
region, the other experimental parameters, 
such as the laser power, eventual pulsing, and detector performance have to be
designed in such a way as to be able to test a photon regeneration probability in the
$10^{-20}$ range (cf. Fig.~\ref{regprop} (bottom panel)).

\section{Experimental Realization}

We propose a photon-regeneration experiment planned around a spare dipole
of the HERA proton storage ring at the DESY magnet test stand. Both parts 
of the experiment, i.e. axion-like particle production and photon-reconversion have to
be accommodated in one single magnet since the test stand architecture in its present configuration
forbids to place two fully functional magnets in line.

\begin{figure*}[hbtp]
  \begin{center}
    \includegraphics*[width=.9\textwidth]{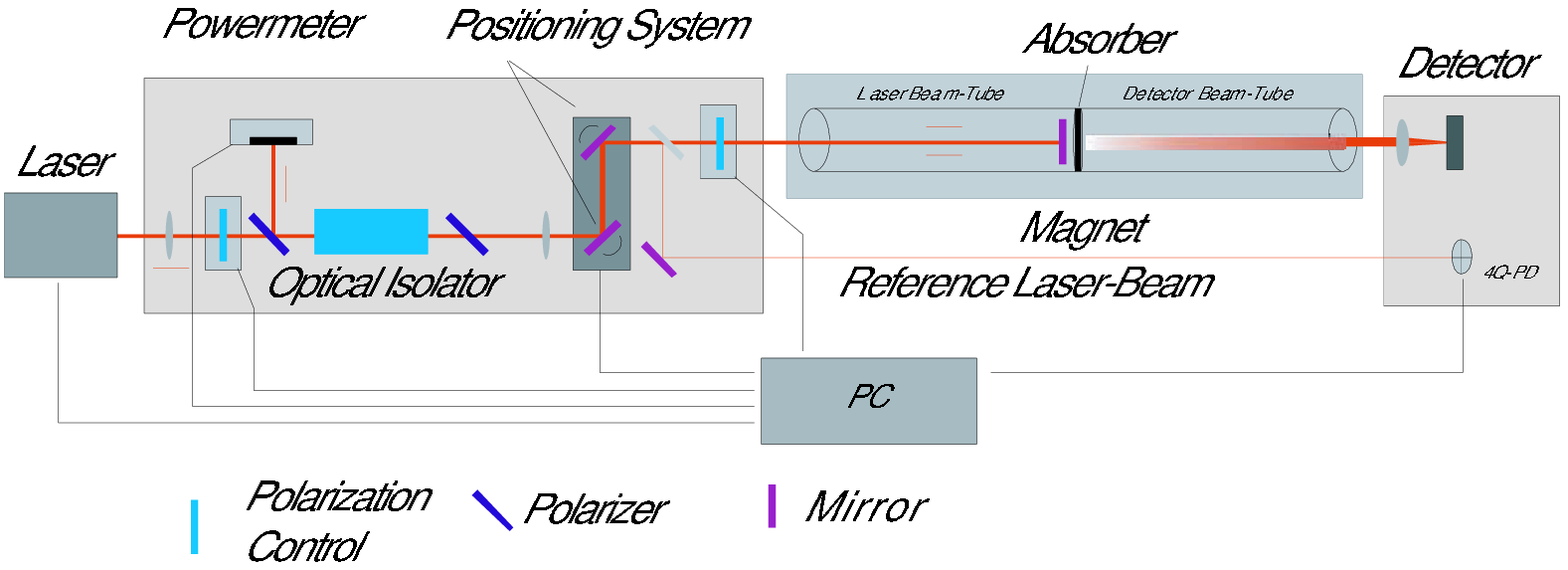}
  \end{center}
  \caption{\label{fig:layout}
    Schematic view of the experimental setup with the laser on the left, 
    followed by the laser injector/extractor system, the magnet and the
    detector table. An intensity-reduced reference beam of the laser is
    guided parallel to the magnet for constant alignment monitoring 
    between laser and detector.
  }
\end{figure*}

The general layout of the experiment is depicted in Fig.~\ref{fig:layout}.
A high intensity laser beam is placed on one side of the magnet traversing 
half of its length. In the middle of the magnet, the laser beam is reflected
back to its entering side, and an optical barrier prevents any photons
from reaching the second half of the magnet. Axion-like particles 
would
penetrate the barrier, eventually reconverting into photons inside the
second half of the magnet. Reconverted photons are then detected with a
pixeled semi-conductor detector outside the magnet. The individual components 
of the experiment are discussed in the following.

\subsection{Magnet}

\begin{figure}
  \begin{center}
    \includegraphics*[width=.45\textwidth]{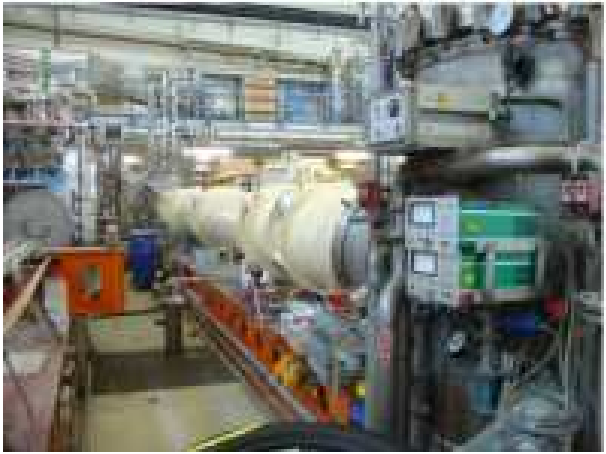} 
  \end{center}
  \caption{\label{heradipole}
    HERA dipole at the magnet test stand.
  }
\end{figure}

The magnetic field will be produced by a spare dipole magnet of the
HERA proton storage ring. At a nominal current of 6000\,A, the magnet reaches
a field of 5.355\,T over a total magnetic length of 8.82\,m. The 
device is in place at the magnet test stand in building 55 
(Fig.~\ref{heradipole}).

The length of the magnet including one compensating end flange is
9766\,mm. The ``open'' magnet is connected at both ends to two
Helium feed boxes of 1180\,mm diameter. These boxes are the interfaces to
the vacuum, to the cold Helium and to the electric current for the
magnet. 

The cold inner beam pipe of the magnet has a temperature of 
4\,K and is bent with a radius of approx. 588\,m to follow the HERA 
curvature. A ``warm''
instrumentation tube of Titanium is inserted into the magnet to allow
for instrumentation for magnetic field measurements. This tube has an
overall length of 13.5\,m, follows the curvature of the beam pipe and acts 
also as a vacuum barrier to the surrounding atmosphere. The instrumentation 
tube is insulated against the cold beam pipe by super insulation. 
During magnet tests, the tube is
flushed with warm Nitrogen to keep it at room temperature. As a side effect 
the heat load on the magnet is high. About 30\,g/s of 4\,K Helium
are needed to keep the magnet cold and to prevent quenches at 6000\,A.

\begin{figure}
  \begin{center}
    \includegraphics*[width=.45\textwidth]{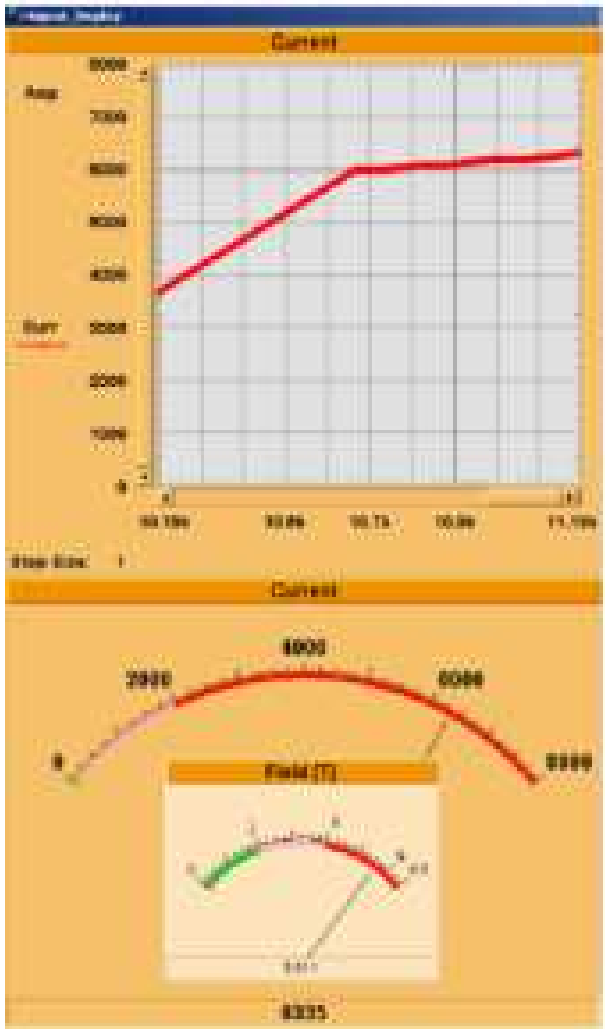} 
  \end{center}
  \caption{\label{magnetrun}
    Screen shot of the magnet control panel from September 26, 2006. At this instance,
    the magnet produced a field of approx. 5.6\,T with a current of about 6300\,A
    (courtesy of H.~Br\"uck).
  }
\end{figure}

After a long shutdown the magnet was successfully put into operation 
again on September 26, 2006. 
A screen shot of the control monitor is shown in Fig.~\ref{magnetrun}.

The position of the Titanium tube was measured by the DESY geometers. 
Although it follows the curvature of the beam pipe,
an open aperture of approximately 18\,mm is available around a straight 
line-of-sight along the whole length, placing geometrical constraints on the
following considerations for the ALPS experiment.

\subsection{Magnet Insert Unit}

The Magnet Insert Unit (MIU) covers all instrumentation inside the magnet. 
It will be inserted into the Titanium tube already installed inside the 
magnet. Following the procedure of standard magnet tests, this tube will be 
flushed with dry nitrogen of around 300\,K temperature while the magnet is 
cooled to liquid helium temperatures.

The main purpose of the MIU is to provide a suitable low pressure environment (LPE) which the laser beam 
passes on its way toward the mirror and back. It needs to 
meet several requirements:
\begin{enumerate}
  \item The space between the MIU and the measuring pipe must allow for a 
    nitrogen flow large enough: to keep the temperature of  the inner 
    measurement-pipe wall everywhere a few degrees above room temperature and 
    to remove the heat produced in the reflecting mirror.
  \item \label{miureq}The pressure of the gas species inside the MIU  is to be varied between 
    1\,$\mu$bar and 10\,mbar. In order to minimize contamination, the pipe has 
    to be evacuated and possibly baked out before introducing the gas. If, at 
    lower working pressures, degassing from the interior of the insert ceases 
    to be negligible,  continuous pumping and replenishing the gas will be 
    required.  In that case, in order to minimize pressure deviations along the 
    length of the magnet, a large diameter of the tube is desirable. Also, the 
    gas may be admitted through a long capillary with several holes along its 
    length. 
  \item The gas pressures in front of and behind the mirror need to be equal. A 
    convenient way to achieve this is to connect both parts by a bypass line. 
    Since absolutely no light from the laser may  propagate into the tube part 
    with the detector, the connection must contain a highly effective light trap. 
    To make the connection symmetrical with respect of flow conductance and to 
    separate gas inlet and vacuum outlet, two pairs of light traps are needed. 
    If the two parts cannot be connected the 
    pressures in both parts need to be measured and controlled separately to make 
    them equal. This method is experimentally more demanding, but does not require 
    light traps.
\end{enumerate}

\begin{figure}
  \begin{center}
    \includegraphics*[width=.45\textwidth]{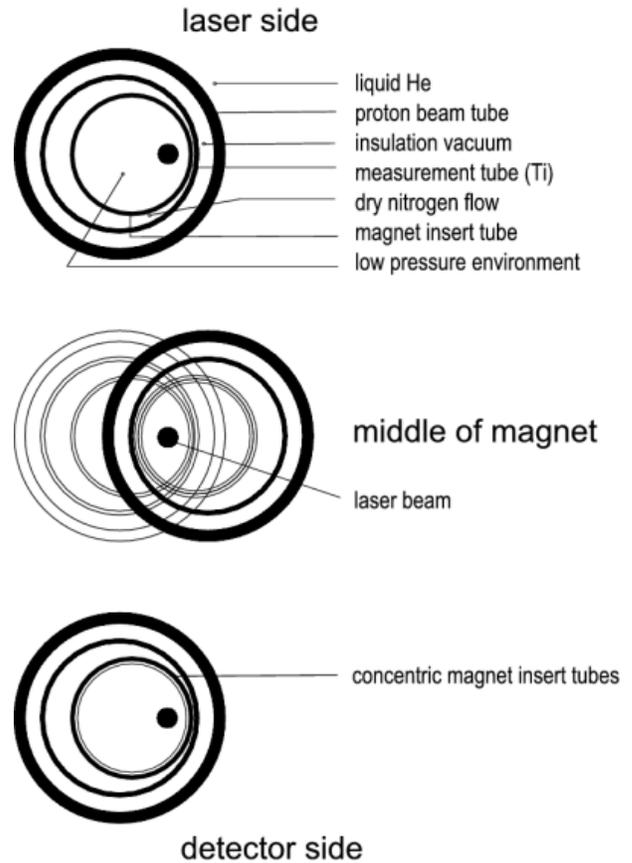}
  \end{center}
  \caption{\label{fig:insert-r-phi}
    View of the magnet insert along the beam line at the beam entrance, 
    at the middle of the magnet, and at the beam exit (top to bottom)
    as seen from the detector.
    The magnet insert cannot fully compensate the curvature of the 
    Ti tube.
  }
\end{figure}

\begin{figure*}
  \begin{center}
    \includegraphics*[width=.9\textwidth]{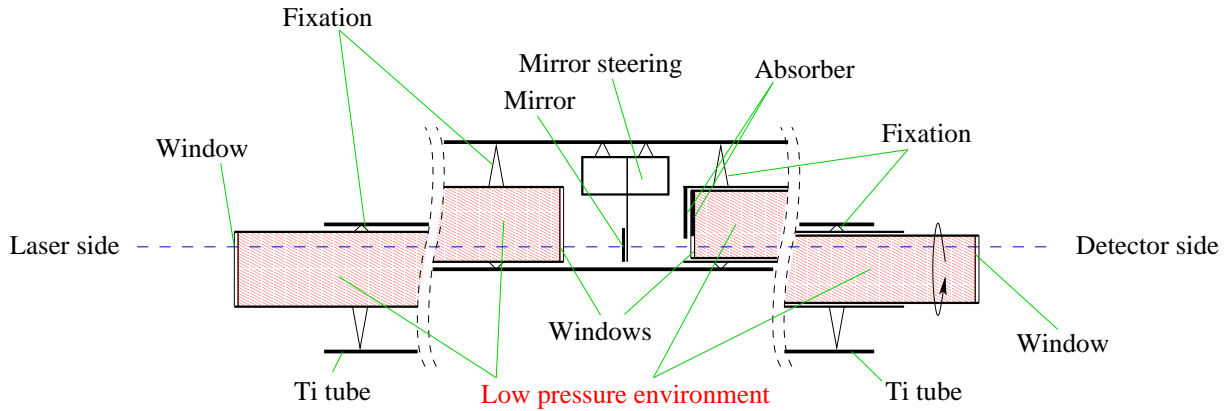}
  \end{center}
  \caption{\label{fig:insert-z}
    Schematic view of the magnet insert perpendicular to the laser beam (dashed line).
    Shown is the installation at the beam entrance, in the middle of the magnet, and at
    the beam exit. Long uniform passages are cut out. Drawing is not to scale.
  }
\end{figure*}

In designing the MIU, a compromise between two diverging objectives has to be found:
reliable, repeatable and easy alignment of the axis of the laser beam and the photon 
detector and complete optical separation of the laser and the detector part, so that 
none of photons produced on the laser side reaches the detector.

We propose to start with an MIU shown in Figs.~\ref{fig:insert-r-phi} 
and~\ref{fig:insert-z} consisting of three parts:
\begin{enumerate}
  \item The laser side part, consisting of a metal tube of 7\,m length, approx. 
    35\,mm outer diameter and 1\,mm wall thickness, with an optical window at each 
    end. It is inserted from the laser side.
  \item A dielectric mirror that reflects the laser beam. It is mounted on a holder 
    such that it can be electro mechanically tilted around two axes which are 
    perpendicular to each other and located in the mirror plane. It is inserted into 
    the measurement pipe either by a special tool or in connection with the laser part. 
    The temperature of the mirror is measured with a temperature sensor. The laser beam 
    is reflected back out of the magnet, separated by the optical insulator and directed 
    into a laser beam dump.  
  \item The detector side part consists of two concentric metal tubes of 1\,mm wall 
    thickness each. The outer tube has an outer diameter of approx. 37\,mm. It is 
    inserted from the detector side. The inner tube is 
    terminated with a full window at the detector end. At the end pointing to the laser there is a 
    smaller, eccentric window and, diametrically across it, a light trap. The inner 
    tube may be rotated between one position where the axis of the laser beam passes 
    the two windows, and another, where the laser beam axis is blocked by the light 
    trap.
\end{enumerate}

Both tube subunits are suspended by (asymmetric) distance holders (fixed to the tubes) 
which glide on the inner wall of the measuring pipe when the MIU is being inserted. 
That way, in the middle of the magnet, they touch the inside wall of the measuring pipe 
on the side with the smaller bending radius and on the opposite side (with the larger 
bending radius) at the ends of the magnet. The mirror surface is optically usable approx. 
1\,mm from its edge. This leaves a horizontal clearance of 14\,mm for the laser beam into 
and out of the tube  subunits. The parts outside the magnet of the laser side tube
and of the inner detector side tube are equipped with vacuum tight connectors for pumping, 
gas inlet and electrical wiring. The space between the MIU and the measuring pipe is 
subjected to a dry nitrogen flow. 

The tube material (e.g. aluminum, stainless steel, OHFC copper, Titan) will be selected 
later in the design process.

An alternative MIU design which concentrates on assuring a complete optical separation between laser and 
detector part, has also been considered and is only summarized here:
\begin{itemize}
  \item The reflecting mirror is placed inside the laser side tube. Behind the mirror 
    the tube is closed by a metal plate so that no light can escape. 
  \item On the detector side there is just one closed tube with a window on the outer 
    end. 
  \item Both tubes are joined in the middle of the magnet by a sliding coupling. 
\end{itemize}

With this alternative MIU the optical adjustment requires a complex procedure.
It relies entirely on the 
permanent correspondence between the position and direction of the laser beam axis 
and reference beams derived from it (cf. Fig.~\ref{fig:layout}). 
Both the detector and the laser assembly need
to be moved in and out of the magnet beam path in a reproducible way. 

In spite of its drawbacks, this second MIU design can serve as a fall-back solution 
in case 
of unsurmountable difficulties with the preferred design as well as a cross checking 
option in case of evidence for axion-like particles. In setting up the experiment, we intend 
to make the necessary preparations to be able to implement and use this MIU design if 
needed.

\subsection{Laser}

\begin{table*}[t!]
  \begin{tabular}{l@{\qquad}c@{\qquad}c@{\qquad}c@{\qquad}c}
    \hline
    Laser Concept & Output Power (W) & Beam Quality ($M^2$) & Polarization & Operation \\
    \hline
    Rod-Laser & 100 - 4.000 & 35 - 75 & random & cw \\
    Disk-Laser & 250 - 8.000 & 12 - 24 & random / linear & cw \\
    Fiber-Laser & 100 - 10.000 & 1.15 - 35 & random / linear & cw, qcw \\
    \hline
  \end{tabular}
  \caption{\label{tab:laser}
    Specification for different commercial available laser systems.
  }
\end{table*}

The design of the laser system is driven by two main aspects: the need of large intensity
of order $10^{21}$\,photons/s and the small aperture of the MIU. In addition, 
controllable linear 
polarization is desirable to distinguish scalar and pseudo scalar axion-like particles 
and to allow for 
systematic tests. Commercial laser 
systems with the required intensity are available for near-infrared wavelengths of order
1100\,nm. Typical parameters for different technologies are listed in Tab.~\ref{tab:laser}.

\begin{figure}[hbtp]
  \begin{center}
    \includegraphics*[width=.45\textwidth]{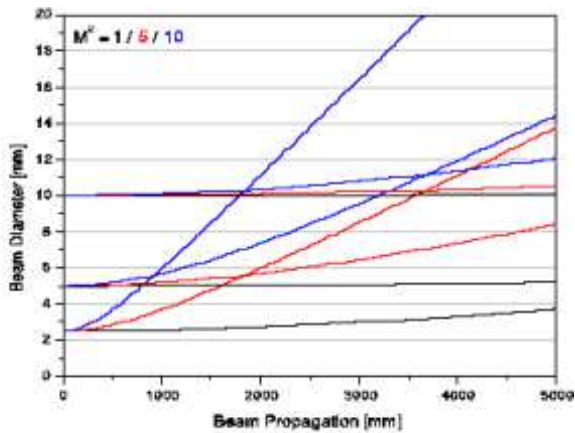}
  \end{center}
  \caption{\label{fig:laserquality}
    Beam propagation within the length of the laser tube for different beam quality parameters and beam 
diameters.
  }
\end{figure}

An important characteristics is the beam quality factor given by
\begin{equation}
  M^2 = \frac{\omega_0 \theta}{\lambda/\pi},
\end{equation}
which depends on the initial beam radius $\omega_0$, the divergence angle $\theta$
and the wavelength $\lambda$. In the proposed setup, the laser beam diameter is
at best equal at the entrance and exit of the magnet, and due to finite divergence 
hence has to be minimal on the mirror in the middle of the magnet. 
Fig.~\ref{fig:laserquality} shows the evolution of beam diameters with the propagation
length for different $M^2$ values and initial diameters. It is evident that beam qualities with 
$M^2<5$ are needed to safely guide the laser through the magnet.

\begin{figure}[hbtp]
  \begin{center}
    \includegraphics*[width=.45\textwidth]{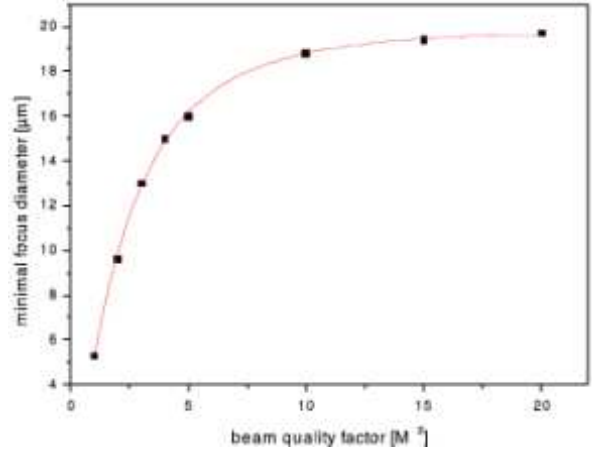}
  \end{center}
  \caption{\label{fig:laserfocus}
    Minimal focus diameter versus beam quality factor $M^2$ for
    a final focus lens with $f=20$\,mm and a beam diameter of
    5\,mm.
  }
\end{figure}

It is planned to focus the reconverted photons on a small region of the 
detector to allow the use of small pixeled semiconductors with low noise. 
The path of incoming laser photons, created axion-like particles and reconverted 
photons lay on top of each other for 
all practical purposes, and so the initial laser beam quality also determines the 
minimal spot size of reconverted photons reachable on the detector. 
A corresponding curve is shown in Fig.~\ref{fig:laserfocus}. The required beam quality can be easily achieved.

In summary: Commercial available laser systems in the near-infrared region with output power above 200\,W 
(corresponds to a flux exceeding $10^{21}$ photons per second) and high beam quality match the requirements 
of the ALPS proposal. However such systems are only available operating in cw-mode.

\subsection{Detector}

Considering the conversion 
probability $P_{\gamma\to\phi\to\gamma} = 10^{-20}$ (Fig.~\ref{regprop}) and the laser properties 
discussed above we expect a signal of the 
order of 10\,photons/s, equivalent to $2\times10^{-18}$\,W. The detector must be able to detect photons 
at this low rate. Although this is challenging, comparable sensitivities are reached in detectors 
for infrared astronomy. 

\begin{figure*}[t!]
  \begin{center}
    \includegraphics*[width=.75\textwidth]{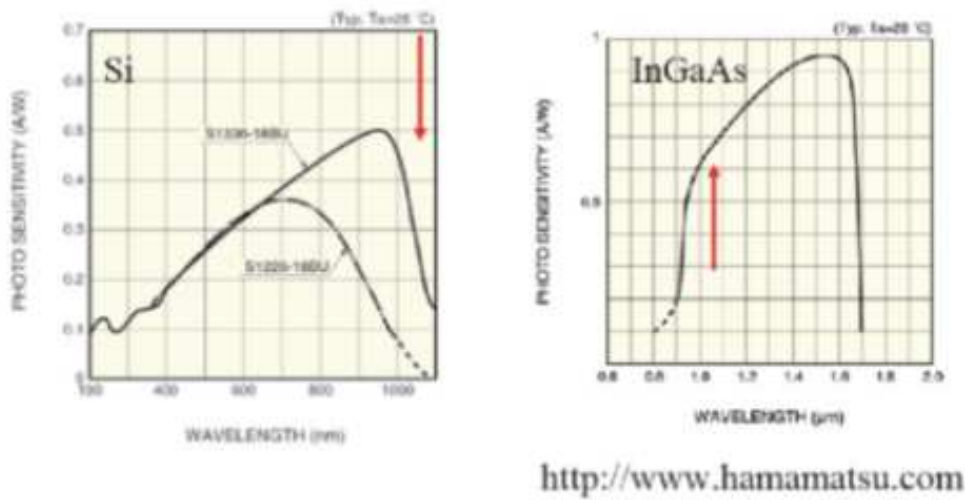}
  \end{center}
  \caption{\label{fig:quantumeff}
    Exemplary sensitivities of Si and InGaAs diodes.
  }
\end{figure*}

In the following we will concentrate on semiconductor detectors. As the ALPS time schedule is somewhat 
ambitious we prefer commercial available systems and can not afford for time consuming extensive R\&D.

Silicon, InGaAs, and HgCdTe have been considered up to now. For these materials 
sensors from single diodes up to two-dimensional cameras are available.
\begin{itemize}
\item Silicon:\\
Silicon detectors normally reach their highest efficiency at wavelengths around 700\,nm. At 1070\,nm the 
quantum efficiencies reach typically only very few percent. Therefore silicon detectors seem to be inadequate 
to our requirements.
\item InGaAs: \\
InGaAs detectors are frequently used in infrared astronomy. Usually they are sensitive to photon wavelengths 
above 800\,nm and up to 1700\,nm (or 2,500\,nm in extended versions).
\item HgCdTe: \\
Such detectors have also been used in infrared astronomy since quite some time. They offer a similar 
sensitivity range as InGaAs detectors.
\end{itemize}

Figure~\ref{fig:quantumeff} shows typical sensitivities of Si and InGaAs diodes.

We have decided to focus on InGaAs in the following while still considering HgCdTe as a possible alternative.

InGaAs sensors are available as single diodes, one-dimensional and two-dimensional arrays. As the laser will 
be essentially operated in cw-mode we do not need fast triggering and gating. Therefore we principally prefer 
two-dimensional arrays:
\begin{itemize}
\item 
They are less sensitive to potential misalignments.
\item
The beam could be focused on the most sensitive pixel.
\item
Pixels outside the signal region could be used to monitor the background conditions.
\end{itemize}

Semiconductor detectors suffer from dark current and read out noise. To reach the required sensitivity the 
total background has to be limited. As an exemplary calculation the maximal background is estimated requiring 
a five standard deviation detection of the signal assuming a measurement with signal $(S+B_1)$ and subtracting 
a background measurement $B_2$.
\begin{equation}
  \frac{(S+B_1)-B_2}{\sqrt{S+B_1+B_2}} > 5
\end{equation}

Assuming $B_1=B_2=B$ and $S = 6,000$ for an integration time of 10 minutes a background rate of about 1 kHz 
could be allowed for considering only statistics. In the experimental set-up it remains to be proven that 
the systematic uncertainties are sufficiently well under control.

The dark current is strongly influenced by the operation temperature. To minimize the technical efforts we 
will consider only cooling by Peltier modules (down to 180\,K) or liquid nitrogen (77\,K).

We would like to stress that in the near infrared region considered here the background induced by thermal 
radiation of the environment amounts to at most only few photons per minute (depending on the size of the 
detector) and hence is neglected.

In summary:
\begin{itemize}
\item
The detector should be based on InGaAs or HgCdTe.
\item
Fast triggering and gating is not required.
\item
The dark current should preferably be below 10$^3$ electrons per second.
\item
The detector will be cooled by Peltier modules or liquid nitrogen.
\item
A two-dimensional array is to be preferred with respect to a one dimensional array or single diodes.
\item
Small area detector elements can be used due to the option to focus the laser beam. This will help to reduce 
the noise.
\end{itemize}

At DESY nearly no experience with low intensity infrared detection exists. In order to build up some 
competence the infrared sensitive camera ST-402 ME-C1 (Santa Barbara Instrument Group) has been ordered in 
summer. Although this Si-detector could not be used at ALPS it would be an easy-to-use device to gain some 
insight into eventual experimental difficulties. However due to the new EU regulations (RoHS) on lead-free 
soldering the camera has not arrived yet.

One of us (GW) has an InGaAs camera at hand (XEVA-USB-FPA-640). At present the properties of the camera are 
investigated, but probably the noise level does not match our requirements. Further tests including cooling 
are under way.

Therefore we do not have at present a detector available matching the requirement. Four directions could be 
followed to arrive at an adequate system:
\begin{itemize}
\item
Continue the tests with the camera at hand and check also other commercially available solutions (i.e. from 
Princeton Instruments).
\item
Check the possibility to develop own read-out electronics (based on commercial available ASICS) for InGaAs 
CCD chips.
\item
Develop a detector based on single diodes.
\item
Check whether a system could be copied or even borrowed from an infrared astronomy group.
\end{itemize}

Significant effort is necessary within the next three month to develop a detector system. Dedicated personnel 
resources are necessary. However no show-stoppers have been identified yet so that confidence exists to reach 
the goals. It would be nice if at least two different systems could be built to allow for systematic studies 
at the experiment.

\subsection{Alignment}

The alignment procedure consists of five steps:
\begin{enumerate}
    \item The inner detector tube is rotated into the position where the laser beam passes the two 
       windows of the inner detector side tube (see Fig.~\ref{fig:insert-z}). The laser beam is directed 
toward the mirror and 
       reflected back into the laser beam dump. 
     \item The detector now is aligned w.r.t. the laser beam by using the fraction of 
       laser radiation not being reflected or absorbed by the mirror (less than 1\,\%). During this 
       step, the detector will be protected by additional laser attenuation outside the magnet and 
       decoupled from the detector.
    \item A low intensity reference beam, generated from the main laser beam with a beam splitter,
       is centered onto a position sensor which is fixed with respect to the detector. From then 
       on the position of the reference beam is continuously monitored using the position 
       sensor. An auto alignment system will use this information to maintain a proper alignment of laser 
beam and detector. 
    \item The inner detector side tube is now rotated into a position where the light trap is behind 
       the mirror, so that any light originating from the laser and passing the mirror is blocked. 
     \item The mirror system inside the magnet might have moved during tube rotations and now needs to 
       be re-aligned in order to reflect the laser beam out of the magnet and into the beam dump again. 
       Neither laser nor detector will be affected by this step, thus the alignment between these 
       two units remains unchanged.
\end{enumerate}

Optionally, the adjustment procedure can be repeated between measurement runs. 

\subsection{Safety}

In the experiment we are confronted with the following hazards:
\begin{itemize}
  \item Cryogenic temperatures;
  \item high magnetic fields;
  \item high electric currents;
  \item intense laser beam.
\end{itemize}
The experiment utilizes the magnet test stand in building 55. Safety
concepts for the first three items are established for this area
and will not be repeated here.

During normal operation the main laser beam will be constrained to 
the tube of the HERA magnet without having a chance to escape. 
Also any reference beam aside the magnet will be guided in closed 
tubes. A detailed safety report (``Gefahrenanalyse'') for the laser 
operation will be prepared once a specific laser system has been 
decided on.

Closed areas with fences and interlock doors will be established 
at both sides of the magnet. A simple electrical interlock loop 
will switch off the laser in case of risk like opening doors, 
quench of magnet, opening of the system etc.

\section{Experimental Reach}

In this section we estimate the sensitivity of the experiment w.r.t. the two 
physics goals: exploring the theoretical parameter space favored by the PVLAS 
results and reaching optimal exclusion limits in case of no signal observation.

The search for axion-like signals will be performed on data taken with and without 
the possibility of axion-like particle production, e.g. with the laser blocked 
before or guided through the magnet. Both 
measurements should be performed alternating to minimize any systematic effects. 
With equal integration time of the background and the signal-plus-background
measurements, we will need a measuring time 
\[
t = 25\,\frac{S+2B}{S^2}\,\,\,[\mathrm{s}]
\]
in order to observe a 5\,$\sigma$ signal excess over background
where $S$ and $B$ are the detected signal and background rates, respectively.

With the variable refraction index tunable with the gas pressure inside the LPE, 
we have the probability to adjust the maximum $\gamma-\phi-\gamma$ reconversion 
probability for each mass hypothesis, i.e. we can tune $F$ from 
equation~\ref{formfactor} to be unity. The rate of reconverted photons then is 
completely determined by the parameters of the magnet and laser systems, with
the coupling as sole free parameter.
Assuming a laser with an output power of 200\,W at 1070\,nm sending its beam into 
a HERA dipole magnet we achieve
\[
 N^\mathrm{ALPS}_\gamma = 1.3\times10^{25}\,\cdot\, \varepsilon \,\cdot\,g^4  \,\,\,[\mathrm{Hz}].
\]
The coupling strength $g$ is given in $\rm GeV^{-1}$; 
$\varepsilon$ is the efficiency of the detector and all optical components, 
which is not yet known exactly. Here we conservatively assume $\varepsilon= 50\,\%$ and 
with $g=10^{-6}\ {\rm GeV}^{-1}$ one achieves a signal $S=N^\mathrm{ALPS}_\gamma = 6\,\mathrm{Hz}$.  
Hence with a background rate of 1\,kHz it follows from the equation given above that 
$t\,\approx\,23\min$ are needed to observe a 5\,$\sigma$ signal. This time is to be doubled 
due to the necessity of a background measurement.

The allowed parameter space from the combination of PVLAS and BFRT is constrained to
three regions with masses roughly between 1 and 3\,meV and couplings
above $10^{-6}$. In order 
to scan these mass ranges, we will need less than 100 different configurations for the 
LPE gas pressure, if the separation of the configurations is chosen such that the 
photon reconversion rate with one configuration is still 90\,\% at the mass for which 
the next configuration is optimal. The total measurement time to probe
the  parameter space allowed by the combination of the PVLAS and BFRT results thus is
\[
100\,\mathrm{configurations} \,\cdot\, 46\,\mathrm{min} \approx 80\,\mathrm{h}.
\]

To first approximation, the experimental reach for the axion mass is determined
by the number of pressure configurations, while the sensitivity for the coupling depends on the
integration time at each configuration. The reach in mass (coupling) depends on the square
(on the fourth power) of the total measuring time, making it hard to explore phase space 
regions well beyond the PVLAS/BFRT allowed region. 

To explore masses up to 5\,meV would require 180 LPE configurations, while already 750
configurations are needed to reach 10\,meV. Similarly, with an hour of
integration time per configuration, the coupling can be explored down to values of 
$5\times 10^{-7}$\,GeV$^{-1}$ with a possible improvement by one order of magnitude in the
more optimistic case of only 10\,Hz background rate.\\

Based on these considerations, we are confident that the proposed ALPS experiment is sensitive
enough to explore the parameter space for axion-like particle production allowed from the combination of PVLAS 
and BFRT. The potential to explore regions of the parameter space much further however are
limited and can only be extended moderately with longer integration times and lower background
rates.
Please note that for the standard QCD axion masses above 100\,eV are predicted for coupling strengths 
above $10^{-7}$\,GeV$^{-1}$. 
A QCD axion with a mass around 1\,meV would have a coupling strength around $10^{-13}$\,GeV$^{-1}$.
Corresponding sensitivities are clearly out of reach of ALPS.

\vspace{.5cm}
\subsection*{Acknowledgments}
\vspace{-.4cm}
We would like to thank Markus Ahlers, Martin Br\"auer, Rolf-Dieter Heuer, 
J\"org J\"ackel, and Ulrich K\"otz for valuable input to this project. We 
are also very grateful to Heinrich Br\"uck, Herman Herzog,
Johannes Prenting, Matthias Stolper and the cryogenics group 
MKS for technical support.

\newpage

\subsection*{Web Links}
\vspace{-.4cm}
\noindent ALPS webpage:\\
{\tt
  http://alps.desy.de/\\
}

\noindent Experimental situation:\\
{\tt 
  http://www.desy.de/$\sim$ringwald/axions/axions.html \\
  http://www.inp.demokritos.gr/$\sim$idm2006/\\
  http://www.sns.ias.edu/$\sim$axions/axions.shtml\\
}

\noindent Detectors:\\
{\tt 
  http://www.XenICs.com/\\
  http://www.sbig.com/\\
  http://www.kodak.com\\
  http://www.lasercomponents.com/\\
  http://www.hamamatsu.com/\\
}

\noindent Experimental setup:\\
{\tt 
  http://www.matthiaspospiech.de/studium/vortraege/\\
  http://www.lzh.de/\\
  http://www.chronz.com/home/index.php\\
  http://www.physikinstrumente.de/de/index.php\\
  http://www.feinmess.de/\\
  http://www.piezojena.com/\\
  http://www.vericold.de/\\
}

\vspace{10.5cm}
\begin{figure}
\end{figure}

\end{document}